\documentclass[aps,pra,twocolumn,showpacs,superscriptaddress,eqsecnum]{revtex4}
\usepackage{epsfig}
\usepackage{graphicx}
\usepackage{bm}
\begin{document}

\date{\today}

\title{Coherent states engineering with linear optics: Possible and impossible tasks}

\author{Bing He}  \email{bhe98@earthlink.net}
\affiliation{Department of Physics and Astronomy, Hunter College of the City University of New York, 
 695 Park Avenue, New York, NY 10021}

\author{J\'{a}nos A. Bergou}
\affiliation{Department of Physics and Astronomy, Hunter College of the City University of New York, 
 695 Park Avenue, New York, NY 10021}

\begin{abstract}
The general transformation of the product of coherent states $\prod_{i=1}^N|\alpha_i\rangle$ to the output state $\prod_{i=1}^M|\beta_i\rangle$ ($N=M$ or $N\neq M$), which is realizable with linear optical circuit, is characterized 
with a linear map from the vector $(\alpha^{\ast}_1,\cdots,\alpha^{\ast}_N)$ to $(\beta^{\ast}_1,\cdots,\beta^{\ast}_M)$.  
A correspondence between the transformations of a product of coherent states and those of a single photon state is established with such linear maps. It is convenient to apply this linear transformation method to design any linear optical scheme working with coherent states. The examples include message encoding and quantum database searching. The limitation 
of manipulating entangled coherent states with linear optics is also discussed. 
\end{abstract}


\maketitle

\section{Introduction} \label{section1}

The robustness of photonic states in an environment with decoherence effects makes them suitable to implementing various quantum information protocols. Since the first proposal for an efficient and scalable quantum computation with linear optics \cite{klm}, single photons have been primarily considered as the carrier of information (for an overview see, e.g., \cite{p-m}). Although there have been quaisi-deterministic schemes based on parametric down-conversion to generate single photons, e.g., \cite{p-j-f,j-p-k}, many quantum information protocols working with single photon signals are still experimentally challenging. Using other types of photonic states to encode quantum information, therefore, attracts the interest of researchers.

The encoding of quantum information on continuous variables (CV) of multi-photon fields has advantages in that their setups are less complex. A number of schemes have been proposed for realizing quantum computation in this approach \cite{f-s,lb,BGS,j-k}. To realize a practical implementation of a scheme working with CV photonic states, the use of the hard non-linear interactions should be minimized. A quantum computation scheme based on coherent sates, a typical example of 
CV photonic states, was thus proposed with the hard nonlinear interactions only required for the off-line preparation of resource states \cite{R-G}. In addition to the theoretical research to apply them as the qubits in computation, coherent states are proposed to perform other quantum information processing tasks, e.g., quantum cryptography protocols \cite{van-Enk, a-c-j} and quantum database searching \cite{s-z-p}. These non-computation quantum information schemes are much more feasible to realize, since they apply linear optics only. Recently, the simplest version of unambiguous identification of coherent states, which is possible to be developed to quantum database searching, has been experimentally realized with 
fiber optics \cite{b-c}.

In this paper, we provide a general structure for the transformations of coherent states that are realizable with linear optical circuits. Coherent states are regarded as the antithesis of single photon states. However, as far as the signal processing with linear optics concerned, it can be shown that there is a correspondence between the transformations of 
these two different types of photonic states. Any permissible linear transformation on a single photon signal, $|\psi_{in}\rangle=\sum_{i=1}^{N}c_ia^{\dagger}_i|0\rangle$, be it unitary or non-unitary, can be deterministically realized with the combination of linear optical modules \cite{reck,bhe}. If we process the signals of $|\psi_{in}\rangle=\prod_{i=1}^{N}|\alpha_i\rangle$, the tensor product of coherent states $|\alpha_i\rangle$, with the same setup, the vector $(\alpha^{\ast}_1,\cdots,\alpha^{\ast}_N)$ (the star stands for the complex conjugate) will transform in the same way as $(c_1,\cdots,c_N)$, the coefficient vector of single photon states. This correspondence between the linear transformations of two types of vectors shows that the different tasks respectively working with coherent states and single photon states can be performed by the same setup.

The rest of the paper is organized as follows: in Sec. II, we present the primary results of the general transformation (unitary and non-unitary) of coherent states under the action of a linear optical circuit; we provide two examples, message encoding and database searching based on coherent states, as the application of the general method in Sec. III; the limitation of linear optics in processing entangled coherent states and the generalization of the circuits to include the non-linearities is discussed in Sec. IV; finally, we summarize the important results we obtain in a brief conclusion.

\section{General transformation of coherent states through linear optical circuit} \label{section2}

We start with the simple situations of the input 
\begin{eqnarray}
|\psi_{in}\rangle=|\alpha_1\rangle|\alpha_2\rangle=D(\alpha_1)D(\alpha_2)|0\rangle 
\end{eqnarray}
being sent to a beamsplitter and a phase shifter. $D(\alpha)$ here represents the displacement operator $e^{\alpha a^{\dagger}-\alpha^{\ast}a}$.
With the interaction Hamiltonian
\begin{eqnarray}
H_{BS}=\theta e^{i\varphi}a^{\dagger}_1a_2+\theta e^{-i\varphi}a_1a^{\dagger}_2,
\end{eqnarray}
of the beamsplitter, the creation operators of the input modes are transformed to those of the output modes $b^{\dagger}_1$ and $b^{\dagger}_2$ as a $SU(2)$ map,
\begin{eqnarray}
\left(\begin{array}{c} b^{\dagger}_{1}\\
 b^{\dagger}_{2}\\
\end{array}\right)
&=&\left(\begin{array}{cc} U_{11} & U_{12}\\
U_{21} & U_{22}\\
\end{array}\right)\left(\begin{array}{c} a^{\dagger}_{1}\\
 a^{\dagger}_{2}\\
\end{array}\right)\nonumber\\
&=&\left(\begin{array}{cc} \cos\theta & ie^{-i\varphi}\sin\theta\\
ie^{i\varphi}\sin\theta & \cos\theta\\
\end{array}\right)\left(\begin{array}{c} a^{\dagger}_{1}\\
 a^{\dagger}_{2}\\
\end{array}\right),
\end{eqnarray}
which defines the unitary operation $U_{BS}$ of a beamsplitter \cite{p-m}.
The reflection and transmission coefficients $R$ and $T$ of
the beamsplitter are given in terms of the parameters as $R = \sin^2\theta$ and $T = 1- R = \cos^2\theta$, and the relative phase $\varphi$ ensures that the transformation is unitary.
The input state of the product of two coherent states is therefore transformed as follows:
\begin{widetext}
\begin{eqnarray}
|\psi_{out}\rangle&=&U_{BS}|\psi_{in}\rangle
=U_{BS}D(\alpha_1)D(\alpha_2) U^{\dagger}_{BS}U_{BS}|0\rangle\nonumber\\
&=&e^{\alpha_1 (U^{\ast}_{11}b^{\dagger}_1+U^{\ast}_{21}b^{\dagger}_2)-\alpha^{\ast}_1 (U_{11}b_1+U_{21}b_2)}
e^{\alpha_2 (U^{\ast}_{12}b^{\dagger}_1+U^{\ast}_{22}b^{\dagger}_2)-\alpha^{\ast}_2(U_{12}b_1+U_{22}b_2)}|0\rangle\nonumber\\
&=& e^{(U^{\ast}_{11}\alpha_1+U^{\ast}_{12}\alpha_2)b^{\dagger}_1-(U_{11}\alpha^{\ast}_1+U_{12}\alpha^{\ast}_2)b_1}
e^{(U^{\ast}_{21}\alpha_1+U^{\ast}_{22}\alpha_2 )b^{\dagger}_2-(U_{21}\alpha^{\ast}_1+U_{22}\alpha^{\ast}_2)b_2}|0\rangle,
\end{eqnarray}
\end{widetext}
where we apply Campbell-Baker-Hausdorff formula $e^Ae^B=e^{A+B}e^{\frac{1}{2}[A,B]}$ from the second line, as well as the unitary transformation in Eq. (2.3). If we define 
\begin{eqnarray}
\left(\begin{array}{c} \beta^{\ast}_1\\
\beta^{\ast}_{2}\\
\end{array}\right)
&=&\left(\begin{array}{cc} U_{11} & U_{12}\\
U_{21} & U_{22}\\
\end{array}\right)\left(\begin{array}{c} \alpha^{\ast}_{1}\\
 \alpha^{\ast}_{2}\\
\end{array}\right),
\end{eqnarray}
the output state will be given as
\begin{eqnarray}
|\psi_{out}\rangle=D(\beta_1)D(\beta_2)|0\rangle=|\beta_1\rangle|\beta_2\rangle.
\end{eqnarray}
Eq. (2.5) characterizes the transformation of the product of two coherent states under the action of Eq. (2.3).

Another basic ingredient in a linear optical circuit is a phase shifter with the interaction Hamiltonian $H=\phi~ a^{\dagger}a$. The creation operator $a^{\dagger}$ of an input mode is mapped to $b^{\dagger}=e^{i\phi}a^{\dagger}$ by the phase shifter, and one input coherent state is thus transformed to $|\beta\rangle=|e^{i\phi}\alpha\rangle$. Analogous to 
Eq. (2.5), this relation is also given as 
\begin{eqnarray}
\beta^{\ast}=e^{-i\phi}\alpha^{\ast}.
\end{eqnarray}

\begin{figure}
\includegraphics[width=66truemm]{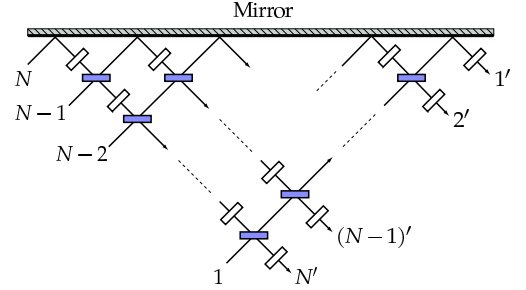}
\caption{(Color online) The unitary transformation module constructed with beam splitters (dark square) and phase shifters (white square). An input of coherent state $|\alpha_1\rangle$ is sent to Port $1$, $|\alpha_2\rangle$ to Port $2$, and so on. The outputs obtained from the ports numbered from $1'$ to $N'$ will be $|\beta_1\rangle$ to $|\beta_N\rangle$ determined by Eq. (2.10). } 
\end{figure}

The unit module of any linear optical circuit is an array of beamsplitters and phase shifters shown in Fig. 1.
Sending a single photon state $\sum^N_{i=1}c_ia^{\dagger}_i$ to this array with the different modes $a^{\dagger}_i$ 
entering the different input ports from $i=1$ to $N$, we will have the following unitary transformation ${\cal U} (N)$ of the mode vector,
\begin{eqnarray}
\left(\begin{array}{c}b^{\dagger}_1\\
b^{\dagger}_2\\ 
\vdots\\
 b^{\dagger}_{N}\\
 \end{array}\right)=\left(\begin{array}{cccc}{\cal U}_{1,1}&{\cal U}_{1,2}&\cdots  &{\cal U}_{1,N}\\
{\cal U}_{2,1} & {\cal U}_{2,2}&\cdots & {\cal U}_{2,N} \\
\vdots&\vdots & \ddots &\vdots \\
{\cal U}_{N,1} &{\cal U}_{N,2}& \cdots & {\cal U}_{N,N}\\
\end{array}\right)\left(\begin{array}{c}a^{\dagger}_1\\
a^{\dagger}_2\\ 
\vdots\\
 a^{\dagger}_{N}\\
 \end{array}\right),
\end{eqnarray}
since it is the product of $SU(2)$ maps $T_{i,j}=T_{i,j}\otimes I_{rest}$ and $U(1)$ maps $e^{i\phi_{i,j}}$ respectively performed by the single beamsplitters and the single phase shifters, which is obtained through the iterative decomposition \cite{reck}
\begin{eqnarray}
{\cal U}(N)T_{N,N-1}\cdots T_{N,1}={\cal U}(N-1)\oplus e^{i\phi}.
\end{eqnarray}
The coefficient vector $\vec {c}=(c_1,\cdots,c_n)$ is correspondingly transformed as $\vec{c'}= \vec{c} ~\cal{U}^{\dagger}$.

Replacing the input with a product of coherent states $|\psi_{in}\rangle=\prod_{i=1}^N|\alpha_i\rangle$, with $|\alpha_1\rangle$ entering the port $1$, $|\alpha_2\rangle$ entering port $2$, and so on, we will obtain the following transformation
\begin{eqnarray}
\left(\begin{array}{c}\beta^{\ast}_1\\
\beta^{\ast}_2\\ 
\vdots\\
 \beta^{\ast}_{N}\\
 \end{array}\right)=\left(\begin{array}{cccc}{\cal U}_{1,1}&{\cal U}_{1,2}&\cdots  &{\cal U}_{1,N}\\
{\cal U}_{2,1} & {\cal U}_{2,2}&\cdots & {\cal U}_{2,N} \\
\vdots&\vdots & \ddots &\vdots \\
{\cal U}_{N,1} &{\cal U}_{N,2}& \cdots & {\cal U}_{N,N}\\
\end{array}\right)\left(\begin{array}{c}\alpha^{\ast}_1\\
\alpha^{\ast}_2\\ 
\vdots\\
 \alpha^{\ast}_{N}\\
 \end{array}\right)
\end{eqnarray}
under the action of the map of the creation operators in Eq. (2.8), which is generalized from that of Eq. (2.5) under the action in Eq. (2.3). If we decompose the above unitary transformation into the product of $SU(2)$ and $U(1)$ maps to make 
the different components of $\alpha^{\ast}_i$ transform as in Eqs. (2.5) and (2.7), combining all these separate maps will give the above transformation of $(\alpha^{\ast}_1,\cdots,\alpha^{\ast}_N)^T$ (T stands for transpose) exactly. 

From this unitary transformation, we also obtain an invariant
\begin{eqnarray}
\sum_{i=1}^N|\alpha_i|^2=\sum_{i=1}^N|\beta_i|^2.
\end{eqnarray}
It is the conservation of the average total photon number $\langle\psi_{in}|\sum_i\hat{n}_i|\psi_{in}\rangle$ under the unitary transformation induced by linear optical circuit. On the other hand, for two sets of coherent states $\{|\alpha_1\rangle, \cdots, |\alpha_N\rangle \}$
and $\{|\beta_1\rangle, \cdots, |\beta_N \rangle\}$ satisfying this relation, we can always find a unitary map ${\cal U}$ realizing the transformations between their products. 

Next, we generalize this type of unitary transformations to the non-unitary ones.
In \cite{bhe} we present a method of combining three linear optical modules in Fig. (1) to realize any non-unitary map ${\cal K}$ of contraction ($||{\cal K}||\leq 1$) on the coefficient vector $(c_1,\cdots, c_N)$ of a single photon state. It is effectively performed by the unitary transformation
\begin{eqnarray}
{\cal U}=\left(\begin{array}{cc} {\cal K} &-(I-{\cal K}{\cal K}^{\dagger})^{1/2}  \\
(I-{\cal K}^{\dagger}{\cal K})^{1/2} & {\cal K}^{\dagger}
\end{array}\right)
\end{eqnarray}
in the extended Hilbert space. ${\cal K}$ here can be any $M\times N$ matrix with the non-zero eigenvalues of ${\cal K}{\cal K}^{\dagger}$ or ${\cal K}^{\dagger}{\cal K}$ being less than or equal to $1$. If $M\neq N$, it can be enlarged by the direction sum with a unit matrix to a square matrix. Following Eq. (2.10), we can implement this extended unitary map on an input of coherent states by sending $|\psi_{in}\rangle=\prod_{i=1}^N|\alpha_i\rangle$ to the input ports of the module, with $|\alpha_i\rangle$
entering the port numbered from $i=1$ to $N$, respectively, and the rest $2\times max(M,N)-N$ ports being dark.
A linear optical circuit modules implementing the unitary transformation ${\cal U}\left (2\times max(M,N)\right)$ thus realizes the following non-unitary map
\begin{eqnarray}
\left(\begin{array}{c}\beta^{\ast}_1\\
\beta^{\ast}_2\\ 
\vdots\\
 \beta^{\ast}_{M}\\
 \end{array}\right)=\left(\begin{array}{cccc}{\cal K}_{1,1}&{\cal K}_{1,2}&\cdots  &{\cal K}_{1,N}\\
{\cal K}_{2,1} & {\cal K}_{2,2}&\cdots & {\cal K}_{2,N} \\
\vdots&\vdots & \ddots &\vdots \\
{\cal K}_{M,1} &{\cal K}_{M,2}& \cdots & {\cal K}_{M,N}\\
\end{array}\right)\left(\begin{array}{c}\alpha^{\ast}_1\\
\alpha^{\ast}_2\\ 
\vdots\\
 \alpha^{\ast}_{N}\\
 \end{array}\right)~~~~
\end{eqnarray}
in a subspace. The output $\prod_{i=1}^{M}|\beta_i\rangle$ is obtained from the output ports numbered from $1'$ to $M'$ in Fig. 1, and some other coherent states also output from the remaining ports. 
A contraction map on a product of coherent states can be, therefore, realized with the same linear optical circuit to implement this contraction map on a single photon state. It is the primary result of this paper and, in the next section, 
we will apply some special forms of Eqs. (2.10) and (2.13) in designing the circuits to process the information encoded in coherent states. 

\section{Linear optical schemes processing information encoded in coherent states} \label{section3}

From the discussion in the last section, we see that it is only necessary to find the proper unitary or non-unitary map and the proper input if we want to realize a certain output state in the form of $\prod_i|\beta_i\rangle$. In what follows, 
we present two examples of how to apply this method in the design of lienar optical circuits.

First, we present a setup to generate the states $|\alpha e^{i\frac{2\pi k}{N}}\rangle$, for $k=0,\cdots,N-1$, which were proposed to realize quantum key distribution (QKD) by linear optics \cite{van-Enk}. A simple choice of the input state is $|\sqrt{N}\alpha\rangle$, and the unitary map is that of discrete Fourier transformation (FT). Out of a circuit corresponding to the following discrete FT, 
\begin{eqnarray}
\left(\begin{array}{c}\alpha\\
\alpha e^{\frac{2\pi i}{N}}\\ 
\vdots\\
 \alpha e^{\frac{2\pi i (N-1)}{N}}\\
 \end{array}\right)&=&\left(\begin{array}{cccc} \frac{1}{\sqrt{N}} & \frac{1}{\sqrt{N}}&\cdots  & \frac{1}{\sqrt{N}}\\
\frac{1}{\sqrt{N}}& \frac{\omega_N }{\sqrt{N}}& \cdots & \frac{\omega_N^{N-1}}{\sqrt{N}} \\
\vdots&\vdots & \ddots &\vdots \\
\frac{1}{\sqrt{N}}& \frac{\omega_N^{N-1}}{\sqrt{N}} & \cdots & \frac{\omega_N^{(N-1)^2}}{\sqrt{N}}\\
\end{array}\right)
\left(\begin{array}{c}0\\
\sqrt{N}\alpha\\ 
\vdots\\
 0\\
 \end{array}\right),\nonumber\\
 && \end{eqnarray}
 all desired outputs will be obtained at the respective output ports. The notation $\omega_N=e^{\frac{2\pi i}{N}}$ is used here.

If we send the input state $|\sqrt{N}\alpha\rangle$ to the first port instead, what we realize
 will be a set of identical coherent states $|\alpha\rangle$. Similarly, any coherent $|\xi\alpha\rangle$ with $|\xi|\leq 1$ can be realized with a combination of unitary and non-unitary maps. For example, continuing to process the output of the identical $|\alpha\rangle$ by a circuit corresponding to 
 \begin{eqnarray}
{\cal K}= \left(\begin{array}{cccc} 1 & & & \\
 & \frac{1}{\sqrt{2}}&  &  \\
& & \ddots &\\
&  & & \frac{1}{\sqrt{N}}\\
\end{array}\right),
 \end{eqnarray}
we will have a series of different coherent states $|\frac{\alpha}{\sqrt{L}}\rangle$, for $L=1,\cdots, N$.

Another interesting application is quantum database searching. This problem is formulated as identifying the input data encoded
by a coherent $|\alpha_0\rangle$ to be which one of $\{|\alpha_1\rangle,\cdots,|\alpha_N\rangle\}$ prepared in a database. The identification of the unknown inputs can be realized by a version of quantum states comparison \cite{b-c-j} as in \cite{s-z-p}. An indispensable requirement for a practical database searching setup, however, is that the identified data $|\alpha_0\rangle$ should be preserved in the process, so that it can be used later. 

We demonstrate the principle of quantum database searching with the simplest case of identifying only two states $|\alpha_1\rangle$ and $|\alpha_2\rangle$. The comparison of states can be realized by a map 
\begin{eqnarray}
{\cal K}=c\left(\begin{array}{ccc} 0 & 0 & 0\\ 
1 & -1&0\\
1 &0 &-1 \end{array}\right),
\end{eqnarray}
which acts on the input $(\alpha_0^{\ast}, \alpha_1^{\ast}, \alpha_2^{\ast})^T$. 
The largest parameter, $c_{max}=\frac{1}{\sqrt{3}}$, is determined by making the eigenvalues of ${\cal K}{\cal K}^{\dagger}$ or ${\cal K}^{\dagger}{\cal K}$ less than or equal to $1$.
We then apply the following unitary map for the identification of $|\alpha_0\rangle$:
\begin{widetext}
\begin{eqnarray}
\vec{\beta}^{\ast}={\cal U}\vec{\alpha}^{\ast}=\left(\begin{array}{cccccc} 0 & 0 & 0 & 1 & 0 & 0\\
\sqrt{\frac{1}{3}} & -\sqrt{\frac{1}{3}}&0 & 0 & \sqrt{\frac{1}{6}} & -\sqrt{\frac{1}{6}}\\
\sqrt{\frac{1}{3}} & 0 & -\sqrt{\frac{1}{3}} & 0 &-\sqrt{\frac{1}{6}} & \sqrt{\frac{1}{6}} \\
\frac{1}{3} & \frac{1}{3}&\frac{1}{3}&0& \sqrt{\frac{1}{3}}&\sqrt{\frac{1}{3}}\\
\frac{1}{3} &\frac{2+\sqrt{6}}{6} &\frac{2-\sqrt{6}}{6} & 0 & -\sqrt{\frac{1}{3}}& 0\\
\frac{1}{3} & \frac{2-\sqrt{6}}{6} & \frac{2+\sqrt{6}}{6}& 0&0&-\sqrt{\frac{1}{3}}
\end{array}\right)\left(\begin{array}{c}\alpha^{\ast}_0\\
\alpha^{\ast}_1\\ 
\alpha^{\ast}_2\\
 0\\
 0\\
 0
 \end{array}\right) =
\left(\begin{array}{c}0\\
\frac{(\alpha^{\ast}_0-\alpha^{\ast}_1)}{\sqrt{3}}\\ 
\frac{(\alpha^{\ast}_0-\alpha^{\ast}_2)}{\sqrt{3}}\\
\beta^{\ast}_4 \\
 \beta^{\ast}_5 \\
 \beta^{\ast}_6
 \end{array}\right).
\end{eqnarray}
\end{widetext}
If there is at least one photon detected form the third output port of the circuit corresponding to this map, the input $|\alpha_0\rangle$ will be identified to be $|\alpha_1\rangle$; similarly, it will be $|\alpha_2\rangle$ if at least one photon is detected from the second output port. 
Except for the output $|\gamma\rangle=|\frac{\alpha_1-\alpha_2}{\sqrt{3}}\rangle$ or $|\frac{\alpha_2-\alpha_1}{\sqrt{3}}\rangle$
consumed in the detection to identify the unknown data, the remaining 
output states $|\beta_4\rangle$, $|\beta_5\rangle$ and $|\beta_6\rangle$, 
which are generated through a non-unitary map
\begin{eqnarray}
\left(\begin{array}{c}
 \beta^{\ast}_4 \\
 \beta^{\ast}_5\\
 \beta^{\ast}_6
 \end{array}\right)=\left(\begin{array}{ccc} \frac{1}{3} & \frac{1}{3} & \frac{1}{3}\\ 
\frac{1}{3} & \frac{2+\sqrt{6}}{6}&\frac{2-\sqrt{6}}{6}\\
\frac{1}{3} &\frac{2-\sqrt{6}}{6} & \frac{2+\sqrt{6}}{6}\end{array}\right) \left(\begin{array}{c}\alpha^{\ast}_0\\
\alpha^{\ast}_1\\ 
\alpha^{\ast}_2
 \end{array}\right),
 \end{eqnarray}
can be well preserved in delay lines. If we input them, together with another prepared $|\gamma\rangle$, to the same circuit but from the inverse direction, the identified data $|\alpha_0\rangle$, as well as the reference states $|\alpha_1\rangle$, $|\alpha_2\rangle$, will be restored from the initial input ports as the result of the ${\cal U}^{\dagger}$ operation. 
In the case that $|\alpha_0\rangle$ is input with the equal {\it a prior} probabilities as 
$|\alpha_1\rangle$ and $|\alpha_2\rangle$, the success probability of this identification process is
\begin{eqnarray}
P_{succ}=1-e^{-\frac{|\alpha_1-\alpha_2|^2}{3}},
\end{eqnarray}
which is as large as the best probability of identifying $|\alpha_1\rangle$ and $|\alpha_2\rangle$ in \cite{s-z-p}.
Provided that the distance, $|\alpha_1-\alpha_2|$, between two states is large enough,  
a setup like this will realize a near deterministic identification, in which the identified input data and the reference states in the database can be also restored for the further processing and the second round searching.

\begin{figure}
\includegraphics[width=100truemm]{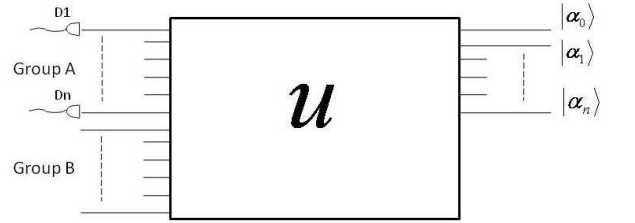}
\caption{The unidentified data $|\alpha_0\rangle$ is sent together with the reference states to the circuit, which implemenets the unitary operation ${\cal U}$. Half of the outputs in group A are measured to determine their identities, and the other half in group B are kept in delay lines. After the measurement results are compared, another set of the determined states are input to the ports of group A, while the outputs of group B are sent back to the circuit. The identified data $|\alpha_0\rangle$ and the reference states are thus retrieved from the initial input ports. Only photodiodes that are unable to resolve photon numbers are required in the detection.} 
\end{figure}

For the general situation of arbitrary number of states, the similar contraction map can be given as
\begin{eqnarray}
{\cal K}=\frac{1}{\sqrt{N+1}}\left(\begin{array}{ccccc} 0 & 0 & 0 &\cdots & 0\\ 
1 & -1& 0 & \cdots & 0\\
1 &0 &-1 & \cdots & 0\\
\vdots & \vdots &\vdots &\ddots &\vdots\\
1 & 0 & 0 &\cdots & -1
\end{array}\right)
\end{eqnarray}
The corresponding setup performs the database searching among the set $\{|\alpha_1\rangle, |\alpha_2\rangle, \cdots, |\alpha_N\rangle\}$.
Two group of outputs are illustrated in Fig. 2: the coherent states in group A will be consumed in the detection for identifying the input data, while those in group B are untouched. The consumed coherent states in group A can be replenished from the extra copies of $|\alpha_i\rangle$ ($i=1,\cdots, N$) processed by the same circuit ahead of the searching.
The maximum number of the required linear optical elements for the circuit grows quadratically with the volume of the database $N$ as $(N+1)(2N+1)$.

Through the above examples, we have demonstrated the power of the general linear map method presented in Section II. To design a circuit that performs a particular task, we only need to determine the required contraction maps ${\cal K}$ or the unitary maps ${\cal U}$, and then we will be able to construct the linear optical circuit following the procedures in \cite{reck} and \cite{bhe}. 

\section{Processing and generating entangled coherent states} \label{section4}
In quantum computation based on the qubit of coherent states $|\alpha\rangle$ and $|-\alpha\rangle$ (see, e.g., \cite{j-k, R-G}), the Bell-cat states (with the global normalization factors neglected)
\begin{eqnarray}
|B_{00}\rangle&\sim&\frac{1}{\sqrt{2}}(|-\alpha,-\alpha\rangle+|\alpha,\alpha\rangle),\nonumber\\
|B_{10}\rangle&\sim&\frac{1}{\sqrt{2}}(|-\alpha,-\alpha\rangle-|\alpha,\alpha\rangle),\nonumber\\
|B_{01}\rangle&\sim&\frac{1}{\sqrt{2}}(|-\alpha,\alpha\rangle+|\alpha,-\alpha\rangle),\nonumber\\
|B_{11}\rangle&\sim&\frac{1}{\sqrt{2}}(|-\alpha,\alpha\rangle-|\alpha,-\alpha\rangle),
\end{eqnarray}
are very useful to the design of various gate circuits. These entangled states are convertible by a 50/50 beamsplitter 
to the cat states $|\sqrt{2}\alpha\rangle\pm |-\sqrt{2}\alpha\rangle$. Although their decoherence times are short for 
the large $|\alpha|$ in lossy optical fiber \cite{a-a-k-j}, the cat states have been found various applications in quantum information processing and metrology (see, e.g., \cite{g-n, g-v}). One question is that if it is possible to produce the Bell-cat states out of a general entangled state
\begin{eqnarray}
|\psi_{in}\rangle\sim\frac{1}{\sqrt{2}}(|\alpha',\beta'\rangle+|\gamma',\delta'\rangle),
\end{eqnarray}
where the parameters $\alpha'$, $\beta'$, $\gamma'$ and $\delta'$ are arbitrary, with only linear optical circuit.
As we have shown in the previous sections, it is to find a contraction map ${\cal K}$ such that
\begin{eqnarray}
\left(\begin{array}{c} -\alpha\\
 -\alpha
 \end{array}\right)&=&{\cal K}\left(\begin{array}{c} \alpha'\\
 \beta'
 \end{array}\right), \nonumber\\
\left(\begin{array}{c} \alpha\\
 \alpha
 \end{array}\right)&=&{\cal K}\left(\begin{array}{c} \gamma'\\
 \delta'
 \end{array}\right),
\end{eqnarray}
if we want to refine the raw entangled state of Eq. (4.2) to, e.g., $|B_{00}\rangle$. By the singular value decomposition ${\cal K}=VDU$, where $U$ and $V$ are unitary and $D$ diagonal, we find the restriction on the parameters of the input states in two situations: (i) if the vectors $v_1=(\alpha',\beta')^T$ and $v_2=(\gamma',\delta')^T$ of the input state are linearly dependent, then the necessary condition to realize the Bell-cat state $|B_{00}\rangle$ is that $v_1=-v_2$;
(ii) if the vectors $v_1$ and $v_2$ of the input state are linearly independent, it is necessary that there exists a unitary map $W$ such that the first or the second component of $Wv_1$ and $Wv_2$ satisfy $(Wv_1)_1=-(Wv_2)_1$ or $(Wv_1)_2=-(Wv_2)_2$, if we intend to have a $|B_{00}\rangle$. Unlike an entangled photon pair like $c_1|HH\rangle+c_2|VV\rangle$ (H and V 
are the polarization modes), which can be always converted by a linear optical circuit to the maximally entangled Bell state $|HH\rangle+|VV\rangle$ with a certain probability, an entangled coherent state will be mapped to a Bell-cat state only conditionally.

There is also a prominent difference between the outputs of coherent-state beams and single photons going through a module 
in Fig. 1. For example, sending single photons to a linear optical module of discrete FT will generate some multipartite entangled states through a post selection \cite{l-b}. However, the output of a product of coherent states is still 
such a product. To realize any type of entangled coherent states, we must add non-linearity to circuit. 
A typical example of such non-linearity is an amplitude-dispersive medium with the Hamiltonian
\begin{eqnarray}
H_{NL}=\hbar \omega a^{\dagger}a+\hbar \chi (a^{\dagger}a)^2,
\end{eqnarray}
where $\omega$ is the frequency of the coherent field and $\chi$ the
strength of the anharmonic term. The input coherent state $|\alpha\rangle$ is mapped to \begin{eqnarray}
K|\alpha\rangle=exp(iH_{NL}t)|\alpha\rangle=\frac{e^{-i\pi/4}}{\sqrt{2}} (|\alpha\rangle+i|-\alpha\rangle)~~
\end{eqnarray}
after an interaction time $t=\pi/\chi$ \cite{j-k}, and an entangled state will be thus obtained with a further action of 
a beamsplitter. Placing these non-linearities between the beamsplitters of a circuit module in Fig. 1, we will generalize its total unitary map in Eqs. (2.9) and (2.10) to
\begin{eqnarray}
{\cal U}=\prod_{i,j} K_{i,j}T_{i,j},
\end{eqnarray}
a product of $K_{i,j}$ induced by the non-linearities and $T_{i,j}$ by the beamplitters (the phase shifter actions are neglected for simplicity). The special examples of this type of setups include the Mach-Zehnder interferometer with Kerr non-linearity in one or two of the arms \cite{s}. With the proper linear and non-linear components in the generalized circuit and the proper input coherent states, we will be able to generate multipartite entangled coherent states of various types.
However, a barrier to constructing such an experimental setup is that the realization of the non-linearity with large 
$\chi$ is still a challenge, if there is no alternative to accomplish the required transformations.

Finally, we provide a general algebraic illustration on the difference between the possible manipulations of entangled photon pairs and entangled coherent-state beams. By introducing a coefficient matrix $\Lambda$, an entangled photon pair $|\Psi\rangle=\sum_{i,j}\Lambda_{i,j}a^{\dagger}_ib^{\dagger}_j|0\rangle$, which occupies
the modes of $a^{\dagger}_i$ for $i=1,\cdots,M$, and of $b^{\dagger}_j$ for $j=1,\cdots, N$, can be given as (the vacuum $|0\rangle$ is neglected) \cite{bhe-07}
\begin{eqnarray}
|\Psi\rangle=(a^{\dagger}_1,\cdots,a^{\dagger}_M)\left(\begin{array}{ccc} \Lambda_{1,1}& \cdots &\Lambda_{1,N} \\
\vdots &\ddots &\vdots \\
\Lambda_{M,1} & \cdots & \Lambda_{M,N}
\end{array}\right)\left(\begin{array}{c}b^{\dagger}_1\\
 \vdots\\
 b^{\dagger}_N\\
 \end{array}\right).\nonumber\\
 &&
\end{eqnarray}
All of its possible transformations by local operations are realizable in terms of the maps on the basis vectors $(a^{\dagger}_1,\cdots,a^{\dagger}_M)^T$ and $(b^{\dagger}_1,\cdots,b^{\dagger}_N)^T$, which can be performed by the respective linear optical circuits of the parties $a$ and $b$ sharing the entanglement. For the entangled coherent-state beams $|\Phi\rangle\sim\sum_{i,j}\Lambda_{i,j}|\alpha_i\rangle_a|\beta_j\rangle_b$ (unnormalized), however, such expression corresponds to a superposition of coherent-state tensor products with $|\alpha_i\rangle$ and $|\beta_j\rangle$ being in more 
than two different locations. The only possibility of evolving the state as two entangled coherent beams on path $a$ and $b$ is to apply the operations on the following superposition of two-dimensional vectors,
\begin{eqnarray}
\Lambda_{11}\left(\begin{array}{c}\alpha_1\\
 \beta_1\end{array}\right)+\Lambda_{12}\left(\begin{array}{c}\alpha_1\\
 \beta_2\end{array}\right)+\cdots+\Lambda_{M,N}\left(\begin{array}{c}\alpha_M\\
 \beta_N
 \end{array}\right),
 \end{eqnarray}
as in Eq. (4.3).

\section{Conclusion} \label{section5}
In conclusion, we present a general method of linear map to study the transformations of coherent states by linear optics. This method is a convenient tool in the design of linear optical circuits to process quantum information encoded in 
coherent states. An example of the applications is a quantum database searching setup based on linear optics, with which the identified input data can be restored for further processing. To generate an entangled coherent state, however, 
it is necessary to introduce non-linearities to optical circuit, and the corresponding linear map should be therefore generalized to include the induced evolutions by the non-linearities.

\begin{acknowledgments}

The authors acknowledge the partial support of grants from PSC-CUNY.

\end{acknowledgments}

\bibliographystyle{unsrt}

\end{document}